# Intelligent Agent for Prediction in E- Negotiation: An Approach

Mohammad Irfan Bala, Sheetal Vij
Department of Computer Engineering
Maharashtra Institute of Technoology
Pune, India
{mirfan508,sheetal.sh}@gmail.com

Debajyoti Mukhopadhyay
Department of Information Technology
Maharashtra Institute of Technology
Pune, India
debajyoti.mukhopadhyay@gmail.com

*Abstract*— With the proliferation of web technologies it becomes more and more important to make the traditional negotiation pricing mechanism automated and intelligent. The behavior of software agents which negotiate on behalf of humans is determined by their tactics in the form of decision functions. Prediction of partner's behavior in negotiation has been an active research direction in recent years as it will improve the utility gain for the adaptive negotiation agent and also achieve the agreement much quicker or look after much higher benefits. In this paper we review the various negotiation methods and the existing architecture. Although negotiation is practically very complex activity to automate without human intervention we have proposed architecture for predicting the opponents behavior which will take into consideration various factors which affect the process of negotiation. The basic concept is that the information about negotiators, their individual actions and dynamics can be used by software agents equipped with adaptive capabilities to learn from past negotiations and assist in selecting appropriate negotiation tactics.

## I. INTRODUCTION

Negotiation is a form of interaction in which a group of agents, with conflicting interests and a desire to cooperate try to come to a mutually acceptable agreement on the division of scarce resources. These resources do not only refer to money but also include other parameters like product quality features, guaranty features, way of payment, etc. Electronic negotiations have gained heightened importance due to the advance of the web and e-commerce. The tremendous successes of online auctions show that the dynamic trade based on e-negotiation will gradually become the core of e-commerce. Whether it is a case of B to B purchase or a case of online shopping [11], it is important to make the traditional negotiation pricing mechanism automated and intelligent. The automation saves human negotiation time and computational agents are sometimes better at finding deals in combinatorally and strategically complex settings.

Traditionally e-negotiation processes have been carried out by humans registering at certain web pages, placing bids, making offers and receiving counter offers from other participants. One major disadvantage with this way of e-negotiation is that the knowledge and experience is kept within the human minds [11]. So humans were replaced by negotiation agents in the process of negotiation. However various problems are faced by the negotiation agents such as limited and uncertain knowledge and conflicting preferences. Also agents may have inconsistent deadline and partial overlaps of zones of acceptance [13]. Moreover, multilateral negotiations are more complicated and time consuming than bilateral negotiations. These factors make it difficult to reach consensus. So decision making mechanism is required to overcome this problem. In Figure 1 we show various modes of interactions in a typical market negotiation framework.

|  |  | Buyer | |
|---|---|---|---|
|  |  | One | Many |
| Seller | One | Negotiation | Auction |
|  | Many | Reverse Auction | Market |

**Fig. 1.** Market negotiation framework

The need is that the agents should be equipped with a decision making mechanism which allows them to adapt to the behavior of the negotiation partner [3]. Intelligent systems for negotiation aim at increasing the negotiators abilities to understand the opponent's needs and limitations. This ability helps to predict the opponent's moves which can be a valuable tool in negotiation tasks. Various negotiation strategies have been proposed which are capable of predicting the opponent's behavior. The research presented here focuses on the online prediction of the other agent's tactic in order to reach better deals in negotiation. While the extensive coverage of all the prediction methods employed in negotiation is beyond the scope of the current work, it is useful to mention several key studies. In this paper we are also proposing a new architecture for prediction of opponent's behavior.

## II. RELATED WORK

Predicting the agent's behavior and using those prediction results to maximize agent's own benefits is one of the crucial issues in the negotiation process. It is necessary for an agent to produce offers based on his own criteria because an agent has limited computational power and incomplete knowledge about opponents. Various approaches [1,2,10,15,16,18] have been proposed for predicting the opponent's negotiation behavior. We reviewed some of the approaches to come up with certain conclusions regarding the efficiency of each approach and their short comings.

Initially game theory was used in the negotiation process. It treats negotiation as a game and the negotiation agents are treated as players of the game. Zeng and Sycara [9] used game-theoretic approach with Bayesian belief revision to model a negotiation counterpart. However game theory has two main drawbacks [1] which make it unsuitable for use in the negotiation process. First is that it assumes the agent has infinite computational power and secondly it assumes all the agents have common knowledge. These limitations of the game theory were overcome by the decision functions. The decision functions produce offers based on the amount of time remaining, resource remaining or the opponent's behavior.

Faratin [18] proposed a bilateral negotiation model in which the two parties negotiate on an issue like price, delivery time, quality etc. The two parties adopt opposite roles (buyer and seller) and use one of the three families of negotiation tactics namely: Time dependant tactics, Resource dependant tactics and behavior dependant tactics. The offers exchanged between the agents are represented as $x_{a \to b}^{t}$. This is the offer generated by agent 'a' for agent 'b' at time 't'. All the offers are restricted in between $min^a$ and $max^a$ which specifies the range of all possible offers of 'a'. Each agent has a scoring function $V^a$ which assigns a score to each offer produced. An agent may respond to the offer by any of the three ways: withdraw, accept or offer

$$response^a(t^n, x_{b \to a}^{t_{n-1}}) = \begin{cases} withdraw(a,b) & if\ t^n > t_{max}^a \\ accept(a,b,x_{b \to a}^{t_{n-1}}) & if\ V^a(x_{b \to a}^{t_{n-1}}) > V^a(x_{a \to b}^{t_n}) \\ offer(a,b,x_{a \to b}^{t_n}) & otherwise \end{cases}$$

$x_{a \to b}^{t_n}$ is the counter offer generated by agent 'a' in response to the offer $x_{b \to a}^{t_{n-1}}$ of agent 'b'. $t_{max}^a$ is the deadline for agent 'a' by which the negotiation should be complete.

Offers generated use one of the three families of tactics [18]. In time dependant tactics time is the predominant factor and each offer generated depends on the amount of time remaining. In resource dependant family of tactics offers depend on how a resource is being consumed. Offers become more and more cooperative as the quantity of the resource diminishes. In behavior dependant family of tactics agent imitates the behavior of the opponent. These tactics differ depending on the behavior of the opponent they imitate and to what degree.

Chongming Hou [1] proposed to use non linear regression approach for the prediction of the opponent's tactics. It could predict the approximate value of opponent's deadline and reservation values. The performance of the agent improved by using this approach as it reduced the number of negotiation breakdowns and caused early termination of unprofitable negotiations. But this approach is restricted for bilateral negotiations only and can be used only when the agent is sure that the opponent is using one of the above mentioned families of tactics for negotiation.

E-negotiation can be classified into three types: one-to-one negotiation, one-to-many negotiation and many-to-many negotiation. Hsin Rau, Chao-Wen Chen, Wei-Jung Shiang and Chiuhsiang Joe Lin [6] focused on one-to-many negotiation architecture and integrated two commonly used coordination strategies i.e. Desperate Strategy and Patient Strategy, to develop a new coordination strategy.

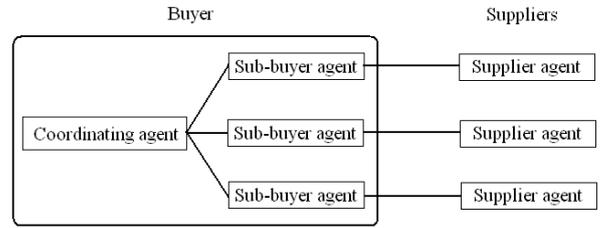

**Fig. 2.** One-to-many negotiation

In one-to-many negotiation (Figure 2), buyer is represented by a combination of one coordinating agent and multiple sub-buyer agents while each supplier is represented by supplier agent. They may use one of the coordination strategies: Desperate strategy and patient strategy. In desperate strategy agents want to complete negotiation process as early as possible. The negotiation process is terminated as soon as any of the sub-buyer agents is successful in negotiation. In case several proposals are found at the same time, the proposal with highest utility gain is accepted. However in patient strategy all the sub-buyer agents are allowed to complete their negotiation process. The sub-buyer agents finishing early are made to wait for other sub-buyer agents till all complete their negotiation. After all the sub-buyer agents finish negotiation, coordination agent selects the best proposal to make the contract. Hsin Rau, Chao-Wen Chen, Wei-Jung Shiang and Chiuhsiang Joe Lin proposed a new strategy called "adapted coordination strategy" [6] which integrates the advantages of above two strategies.

Many other prediction approaches have been proposed which are based on machine learning mechanism. Most of the work devoted to the learning approach is focused on learning from previous offers i.e. offline learning. It includes: Bayesian learning, Q-learning, case-based reasoning and evolutionary computation. They require training data and such agents need to be trained in advance. However this approach may not always work well for the agents whose behavior has been excluded from the training data. Also such data may not be always available. This issue was overcome by Fenghui Ren and Minjie Zhang [5] who proposed three regression functions namely linear, power and quadratic to predict agent's behavior. These regression functions use only data about historical offers in the current negotiation thread instead of the using training data which may not always be available. These three regression function are given below and cover most of the negotiation behaviors of the general agents.

- Linear Regression Function
$$u = b * t + a$$
- Power Regression Function
$$U = a * t^b$$

- Quadratic Regression Function
$$U = a * t^2 + b * t + c$$

where u denotes the estimated value for an agent's utility, $t\ (0 \leq t \leq \tau)$ denotes the negotiation time and a, b and c are the parameters which need to be calculated. Parameters a, b and c are independent of t.

Brzostowski and Kowalczyk [10] presented a way to estimate partners' behaviors by employing a classification method. They used a decision making mechanism which allows agents to mix time-dependant tactics with behavior dependant tactics using weights which can result in quite complex negotiation behavior. However this approach only works for the time dependent agent and the behavior-dependent agent, which limits its application domains. Gal and Pfeffer presented a machine learning approach based on a statistical method [14,17]. The limitation of this approach is the difficulty of training the system perfectly. Therefore, for some unknown kind of agents whose behaviors are excluded in the training data, the prediction result may not reach the acceptable accuracy requirements.

I. Roussaki, I. Papaioannou, M. Anagnostou [13] proposed an approach based on learning technique which has been employed by Client Agents and uses a feed-forward back-propagation neural network with a single output linear neuron and three hidden layer's neurons. These neural networks require minimal computational and storage resources making it ideal for mobile agents. Also the system does not require information about the previous records. Only information about the current negotiation process is taken into consideration. The agents use a fair relative tit-for-tat negotiation strategy and the results obtained were evaluated via numerous experiments under various conditions. The experiments indicated an average increase of 34% in reaching agreements [13]. This approach has excellent performance when the acceptable interval of the negotiation issue overlaps irrespective of the concession rate. On the other hand if the acceptable intervals' overlap is limited and the deadline is quite high, this approach is likely to fail. Also this work was restricted to single issue and bilateral negotiations only.

### III. PROPOSED ARCHITECTURE

We are proposing the architecture of behavior prediction module in the form of web services as depicted in Figure 3. It has already been established in [4] that providing negotiation as a service (NaaS) is a completely innovative application model of software which provides services through internet. Its benefits are , we can obtain stable visiting quantity, user need not concern about maintenance and upgrade of system as it is done on the server independently, saving human and material resources, automated negotiation system can make use of the existing basic facilities provided by e-commerce platform i.e. security, authentication, transaction management etc. ,saving costs of development.

Working: The seller will advertise itself through a well known service registration center which will make it visible to all the interested buyers. All the available services at any point of time are stored in the service registration center. A buyer looking for some product will query the service registration center to discover the product of his interest. Once the preferences are matched, buyer and seller will directly communicate with each other and start negotiation. Each buyer and seller has its own module for behavior prediction. Complexity of the behavior prediction module may vary depending on the number of issues taken into consideration. Also the negotiations may be bilateral or multi lateral which will make the prediction process more complicated. Here we have taken seven issues into consideration during prediction: Original price, age, culture, time, quantity, quality and feedback although the number of issues may increase or decrease with corresponding increase or decrease in the complexity of the behavior prediction module.

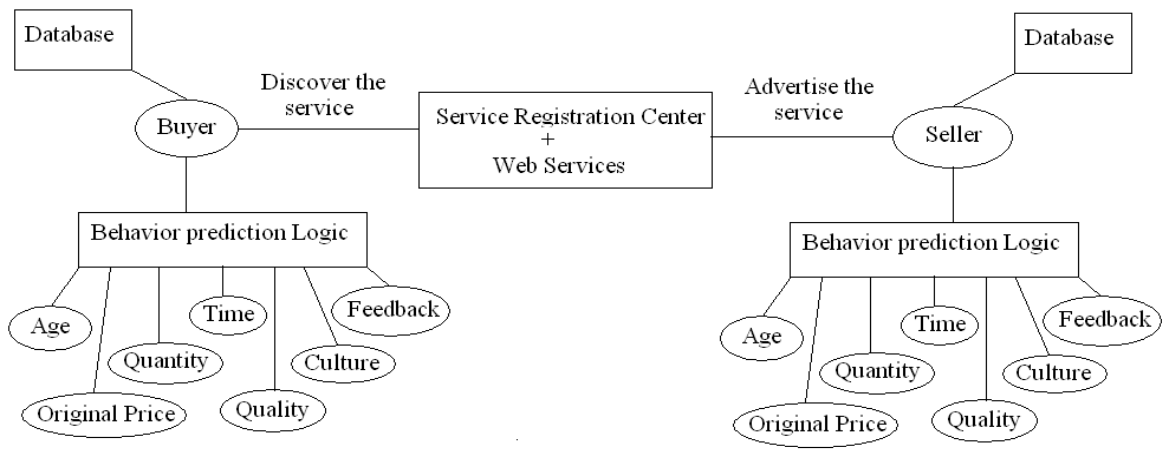

**Fig. 3.** Proposed Architecture for behavior prediction

The behavior prediction logic will use artificial neural networks which have been proved to be universal approximators when provided with sufficient hidden layer neurons and assuming that the activation function is bounded and non-constant. Neural networks also possess the abilities of being self adaptive and self learning. During the first few iterations of negotiation, behavior prediction module will not be used and all the offers and counter offers will be stored in the database and an attempt will be made to try to find the decision function used by the opponent. Later the data stored in the database is used to train the negotiating agent which can predict the opponent's offers.

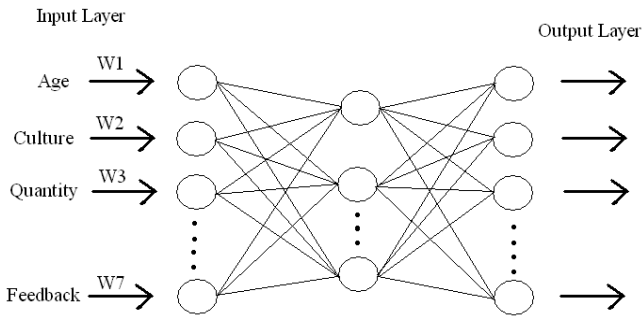

**Fig. 4.** Structure of the artificial neural network

The structure of the artificial neural network used in the proposed system is given. All the issues included in the behavior prediction logic are given as input to the system and each issue is assigned some weight depending on its importance. The input for each neuron is not the input variable or the signal coming from previous neuron but it is the signal multiplied by the weight assigned to that issue. Issues with continuous values should be provided discrete values first to make the process of behavior prediction easy. Example: Instead of using the continuous values for 'age', it should be grouped as youth, middle-aged and old for age group of [10-25], [25-50], [50+] correspondingly.re you begin to format your paper, first write and save the content as a separate text file. Keep your text and graphic files separate until after the text has been formatted and styled. Do not use hard tabs, and limit use of hard returns to only one return at the end of a paragraph. Do not add any kind of pagination anywhere in the paper. Do not number text heads-the template will do that for you.

The architecture shown is for bilateral negotiations. However it can be extended to support multi lateral negotiations where each pair of agents has similar architecture in between them.

*A. Sample Negotiation Scenario*

Consider a sample negotiation scenario where the two companies are trying to secure a contract between them. Suppose company A wants to sell aircrafts and which company B is considering to purchase. The company A will register itself with the well known registration center and advertise itself making it visible to all buyers. Company B will use the registration center to find the company A which can satisfy its requirements. Once the companies meet they will start communication and will first decide the issues of conflict. Here we consider only three issues: Price, quantity and warranty. Both the companies will be using some strategy for negotiation which is private and is not disclosed during the negotiation process. Each company will assign some weight to these issues such that total of all the weights is 100.

| Negotiation issue | Company A weights | Company B weights |
|---|---|---|
| Price | 50 | 60 |
| Quantity | 20 | 15 |
| Warranty | 30 | 25 |

**Table 1**. Relative ratings of the two companies

Each issue has one or more options, for example, price has three options: $1 million, $1.1 million and $1.2 million. Each option should also be rated like the negotiation issues were rated. Total rating of the options should preferable add up to 100 although not necessary. Similarly quantity may have two options: 3 and 5, and warranty may have 4 options: no warranty, 6 months, 1year and 2 years. Each option of each issue should be rated before starting the negotiation process. Also the rating of the least desired option of each issue should have zero rating making it a threshold value beyond which negotiation will terminate. The agent will generate all possible permutations of the offers and calculate the total profit of each offer in advance. None of the given two companies knows about the preference structure of the other company and at no point are these preferences revealed.

Once the assignment of ratings is complete, any of the two companies may start the negotiation process. Whenever a counter offer is received, utility function is used to calculate the concession offered over the previous offer. The utility function is given as:

$$\text{Total profit} = \sum_{i=1}^{n} \text{weight of the } i^{th} \text{ issue} \; \frac{\text{Rating of offered option of } i^{th} \text{ issue}}{\text{maximum rating of } i^{th} \text{ issue}}$$

Where n is the number of issues, 'rating of offered option of $i^{th}$ issue' is the rating of the $i^{th}$ issue in the proposed offer with maximum rating of each issue given on the denominator.

The negotiation process continues with several offers and counter offers. Each offer should provide increased profit than the previous offer. Increase in profit after each negotiation round indicates that the negotiation process is converging and the possibility of negotiation process resulting in agreement increases. In case multiple offers are received with decreasing utility function, the agent may decide to end the negotiation process early to reduce the cost of unsuccessful negotiations. Negotiation can also terminate if the opponent crosses the threshold values for any of the issues.

## IV. CONCLUSION AND FUTURE WORK

This work reviews the various methods used for predicting the opponent's behavior and then proposes architecture for behavior prediction using artificial neural networks. It proposes the use of database for storing the results and suggests various issues that can be taken into consideration while predicting the opponent's behavior. The proposed intelligent agent based architecture is for bilateral negotiations and may be extended in future to multi lateral negotiations. The given architecture is for general use and may not produce optimal results in all situations. So a situation specific architecture is required in every case of negotiation, where the negotiation issues are selected accordingly. In future we would be making the system to simulate above architecture with the application of agent's behavior prediction in web based negotiation. We plan to test it vigorously and do the necessary comparative study and analysis with above mentioned related systems. We can also extend our research in behavior prediction of the agents in the direction of multilateral negotiations after successful completion of bilateral system.